# A physics and learning-based transmission-less attenuation compensation method for SPECT

Zitong Yu[a], Md Ashequr Rahman[a], Thomas Schindler[b], Richard Laforest[b], Abhinav K. Jha[a,b]

[a]Department of Biomedical Engineering, Washington University in St. Louis, St. Louis, MO, USA, 63130;
[b]Mallinckrodt Institute of Radiology, Washington University in St. Louis, St. Louis, MO, USA, 63110.

## ABSTRACT

Attenuation compensation (AC) is a pre-requisite for reliable quantification and beneficial for visual interpretation tasks in single-photon emission computed tomography (SPECT). Typical AC methods require the availability of an attenuation map, which is obtained using a transmission scan, such as a CT scan. This has several disadvantages such as increased radiation dose, higher costs, and possible misalignment between SPECT and CT scans. Also, often a CT scan is unavailable. In this context, we and others are showing that scattered photons in SPECT contain information to estimate the attenuation distribution. To exploit this observation, we propose a physics and learning-based method that uses the SPECT emission data in the photopeak and scatter windows to perform transmission-less AC in SPECT. The proposed method uses data acquired in the scatter window to reconstruct an initial estimate of the attenuation map using a physics-based approach. A convolutional neural network is then trained to segment this initial estimate into different regions. Pre-defined attenuation coefficients are assigned to these regions, yielding the reconstructed attenuation map, which is then used to reconstruct the activity distribution using an ordered subsets expectation maximization (OSEM)-based reconstruction approach. We objectively evaluated the performance of this method using highly realistic simulation studies conducted on the clinically relevant task of detecting perfusion defects in myocardial perfusion SPECT. Our results showed no statistically significant differences between the performance achieved using the proposed method and that with the true attenuation maps. Visually, the images reconstructed using the proposed method looked similar to those with the true attenuation map. Overall, these results provide evidence of the capability of the proposed method to perform transmission-less AC and motivate further evaluation.

## 1. INTRODUCTION

Single-photon emission computed tomography (SPECT) is a widely used clinical nuclear-medicine imaging modality that plays an important role in several diseases, including cardiovascular and neurodegenerative diseases. A major image-degrading process in SPECT is the attenuation of photons as they propagate through the body before reaching the detector. Compensating for attenuation is a pre-requisite for reliable quantification and beneficial for visual interpretation tasks in SPECT[1],[2]. Multiple methods have been proposed for attenuation compensation (AC) in SPECT[3]-[5]. Performing AC requires an attenuation map of the patient. This attenuation map is typically obtained from a transmission scan, such as a CT scan. However, this leads to increased radiation dose, higher costs, and possible misalignment between SPECT and CT scans[3],[6]. Typically, these CT scans are obtained from dual-modality SPECT systems. A major portion of the SPECT market share is occupied by stand-alone SPECT systems[7] that do not have a CT component. Further, several emerging solid-state-detector-based SPECT systems, which provide high quality images at low dose, do not have CT-imaging

---

capability[8]. Finally, mobile SPECT systems that enable imaging in remote locations, are typically SPECT only. Therefore, methods that can perform AC without requiring a transmission scan are much needed[1].

Research in transmission-less AC in SPECT has had a rich history. Methods have been proposed that estimate the attenuation coefficients directly from the photopeak window of the SPECT emission data. These methods either apply iterative inversion [9]-[12] or exploit the consistency conditions in the forward model.[13]-[15] However, these methods are slow and have met with limited success[1]. More recent studies have shown that scattered photons in SPECT contain information to estimate the attenuation distribution[17]. Additionally, methods that estimate the attenuation map from the scatter-window data have shown promise [18-23]. Based on this premise, our first objective in this manuscript is to propose a physics and deep-learning-based three-dimensional (3D) reconstruction method that uses the SPECT emission data in the photopeak and scatter windows to perform transmission-less AC.

The second objective of this manuscript is to objectively evaluate the proposed method on the clinical task of detecting cardiac defects in myocardial perfusion SPECT images using a virtual imaging trial-based framework. Our studies show that the proposed method yields equivalent performance on the task of detecting cardiac defects in myocardial perfusion SPECT images as that obtained when the attenuation map is known. In the next section, we first provide a brief description of the proposed method.

## 2. METHODS

### 2.1. The proposed method

An overall schematic of the proposed method is shown in Fig. 1. The proposed method uses the scatter-window data to estimate the attenuation map, which is then used to estimate the activity map from the photopeak data.

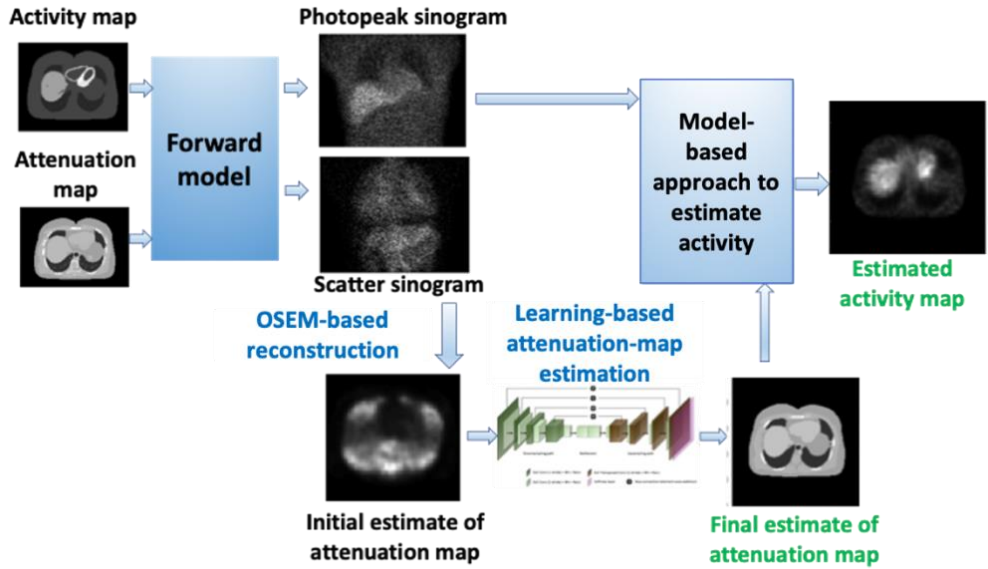

Fig. 1: The workflow of the proposed method

Attenuation in SPECT is primarily due to Compton scattering. The probability of Compton scatter at a given location is directly proportional to the attenuation coefficient at that location. Thus, reconstruction with the scattered photons may serve as a source of contrast between tissues with different attenuation coefficients. This idea has been previously explored in several studies, where the scatter-window data was used to reconstruct the attenuation map using filtered back-projection (FBP)-based approaches[20]-[23]. However, FBP-based approaches have limited ability to model noise and system physics. We thus used an ordered-subsets expectation maximization (OSEM) approach for this purpose, with the goal of accounting for the photon noise and the imaging-system instrumentation, including the collimator and the detector configuration. This yielded an initial 3D estimate of the attenuation map. We observed that this initial estimate, while it provides some contrast, visually looked different from the true attenuation map, as we see in Fig. 1.

Our next goal was to segment this initial estimate of the attenuation map into different regions. For this purpose, we used a deep-learning-based approach. Deep learning has shown significant promise in image segmentation[24]-[26]. We trained a convolutional neural network (CNN) to segment the initial estimate of the attenuation map into regions with different attenuation coefficients. A schematic of the architecture of the CNN is shown in Fig. 2. The architecture was similar to a U-net, with modifications made to optimize for this segmentation task. During the training process, the CNN was provided the initial 3D estimate of the attenuation map as the input, and the corresponding true or CT-based attenuation map that had been segmented into regions with different attenuation coefficients as the output. The cross-entropy loss function between the estimated and true segmented attenuation maps was minimized. Once trained, the CNN, when input an initial estimate of the attenuation map, outputs the segmented attenuation map. Pre-defined attenuation coefficients were then assigned to each region in this segmented map, yielding the final estimate of attenuation map.

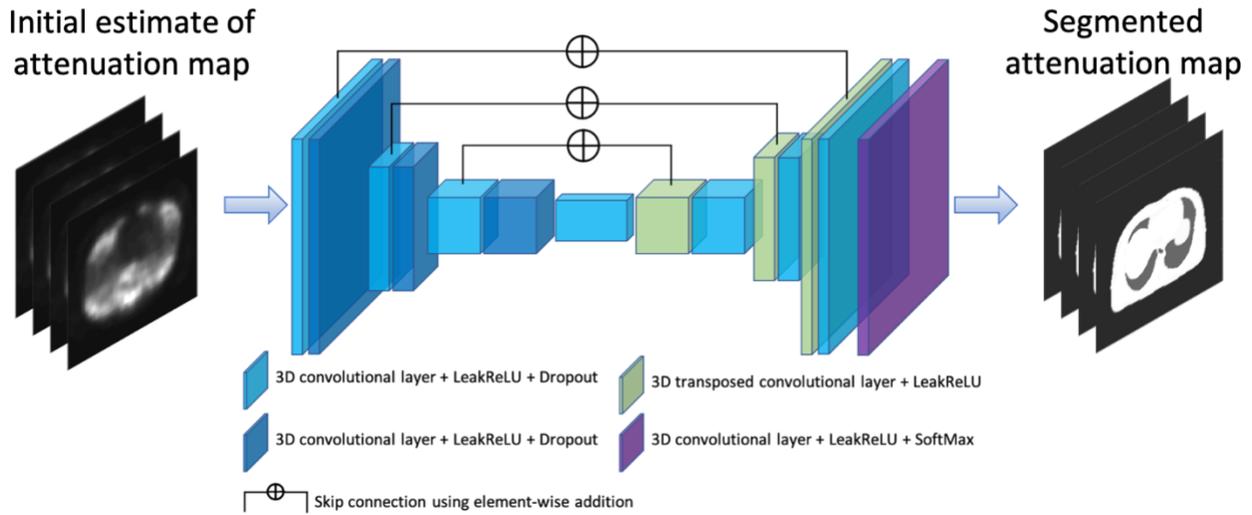

Fig. 2: Schematic of the proposed CNN-based approach to segment the initial estimate of the attenuation map

Using this estimate of the attenuation map and the photopeak projection data, we reconstructed the activity map using a model-based reconstruction approach. This reconstruction approach was based on an OSEM-based algorithm that compensated for attenuation, scatter, collimator-detector response, and noise. Thus, the proposed approach yielded the reconstructed activity images using only the SPECT emission data.

**2.2.    Objective evaluation of the proposed method**

SPECT images are acquired to perform clinical tasks, such as detection, quantification, or a combination of both. Our method incorporates a deep-learning-based component. Conventionally, deep-learning-based methods have been evaluated using figures of merit (FoMs) such as root mean squared error, structural similarity index, and peak signal-to-noise ratio. However, it was shown in a previous study that evaluation using these metrics may not correlate with performance on detection tasks in myocardial perfusion SPECT[30]. More specifically, in that study, a deep-learning-based method for denoising low-dose myocardial perfusion SPECT images was evaluated using both conventional metrics and on the clinical task of defect detection. It was observed that evaluation with conventional metrics suggested that the deep-learning-based method was superior to a simple physics-based approach. However, when the approach was evaluated on the detection task, it was observed that there was no difference in performance between the deep-learning and the physics-based methods. Thus, in this study, we objectively evaluated our method on the clinical task that was required from the image. The study was conducted in the context of evaluating the proposed method on the task of detecting cardiac defects in myocardial perfusion SPECT.

Conducting this evaluation study required knowledge of the ground truth. For this purpose, we used a virtual imaging trial (VIT) framework. As has been shown in several studies, including those evaluating SPECT methods[27]-[31], the VIT framework evaluates the performance of a method against known ground truth by providing the ability to accurately model *in vivo* anatomical and physiological properties and imaging system characteristics, incorporate population variability, and generate multiple scan realizations of the same patient to evaluate reproducibility. Even more importantly, this is all done *in silico*, which is inexpensive and enables optimizing the method before conducting clinical studies. However, for these studies to be clinically relevant, it is important that they are clinically realistic. Below, we describe the process to conduct this study to ensure clinical realism.

### 2.2.1. Virtual patient population

A virtual population of patients imaged with myocardial perfusion SPECT was generated (N = 1600). In this population, variability in both patient anatomy and physiology, as well as variability in the defects, was modeled. Anatomical templates for these phantoms were obtained using the 3-D extended cardiac and torso (XCAT) phantom, with variation in body and heart sizes sampled from distributions in a previously conducted study[32]. Both male and female patients were simulated. In this population, half of the phantoms contained a cardiac defect. We simulated four different types of cardiac defects, which varied in the locations, severities, and shapes. Next, for each anatomical template, we used activity distributions from clinical Tc-99m-sestamibi MPS studies to determine activity uptake in the various organs[33]. We generated the true attenuation maps corresponding to these phantoms with attenuation coefficient values at 140 keV. Further, the true attenuation maps were segmented into seven regions, namely lung, soft tissue, muscle, spine, ribs, lung nodules, and background, yielding the true segmented attenuation maps. Thus, we had a digital phantom population of 1600 patients with anatomical templates and defect types based on population data and tracer distributions guided by clinical data. Further, for these simulated patients, the true segmented attenuation maps were also available.

### 2.2.2. System simulation and image reconstruction

A clinical 3D SPECT system with parameters similar to a GE Discovery 670 (GE Healthcare, Haifa, Israel) with a parallel-hole low energy high-resolution (LEHR) collimator was simulated. The system had an energy resolution of 9.8% at 140 keV, intrinsic resolution 0.39 cm at 140 keV, and 9.5 mm crystal thickness. Projection data in both the photopeak (126-154 KeV) and the scatter window (90-122 KeV) were obtained using SIMIND[34], a validated Monte Carlo-based SPECT simulation software. The simulation accurately modeled the various image-degrading processes in SPECT including scatter, attenuation, photon noise, collimator response, finite energy, and spatial resolution of detector. The SPECT projections were obtained at 60 angles over 180 degrees orbits.

The initial estimate of attenuation map was reconstructed from the scatter-window data using the OSEM-based reconstruction approach with four iterations and four subsets. This yielded 1600 initial estimates of the attenuation maps. Of these, 800 attenuation maps, along with the corresponding true segmented attenuation maps, were used to train the CNN on the task of segmentation. For this purpose, the autoencoder was provided the true segmented attenuation maps and was trained by minimizing the cross entropy between the true and estimated segmented attenuation maps. The CNN, as shown in Fig. 2, was trained using the Adam optimizer[35]. We implemented our network using TensorFlow 1.10.0 with Keras on NVIDIA Titan RTX GPU with 24 GB of memory.

Pre-defined attenuation coefficients were assigned to segmented region, yielding the final estimates of attenuation maps. The activity images were reconstructed using the OSEM-based approach with four iterations and four subsets.

### 2.2.3. Observer study

The performance of the proposed method was objectively evaluated on the task of detecting defects from the reconstructed activity images, which is a main clinical task for which myocardial perfusion SPECT images are acquired. In our study, defects varied in location, shape, severity, and the rest of the body varied in size and activity uptakes. Thus, the variations were present in both the signal and the background. This is referred to as a signal-known statistically/background-known statistically (SKS/BKS) task. To conduct the observer study, we used a previously validated model observer designed precisely for this task[28]. To implement this observer, we divided the data into four sub-ensembles according to the defect types. The defect-free images, which were defined as healthy images, were randomly assigned into these four sub-ensembles uniformly. Images having the centroid of the defects at the center of the images, were extracted to a size of $32 \times 32 \times 32$. For a sub-ensemble $j$, denote the mean data vector by $\bar{g}_j$, and denote the covariance matrix of the image data by $K_{gj}$. We used five 3-D LG channels operating on the image vector $g_j$. Let $U$ denote the channel matrix and $v_j = U^T g_j$. The test statistics were calculated as

$$\lambda(Ug_j) = \lambda(v_j) = \Delta \bar{v}_j^T K_{vj}^{-1} v_j + \frac{1}{2}\left(\bar{v}_{j0}^T K_{vj}^{-1} \bar{v}_{j0} - \bar{v}_{j1}^T K_{vj}^{-1} \bar{v}_{j1}\right).$$

From these test statistics, the receiver operating characteristics (ROC) curves were plotted and the area under the ROC curve (AUC) was calculated. LABROC4 was used to estimate ROC curves [36].

As mentioned above, the proposed method was evaluated on 800 test images. The method was compared with two other methods: (a) The reconstruction obtained when the true attenuation map was used. We refer to this as the true attenuation map-based AC (TAAC) method. (b) The reconstruction method that used a uniform attenuation map. We refer to this as the uniform attenuation map-based AC (UAAC) method.

## 3. RESULTS

### 3.1 Objective evaluation

The ROC curves obtained by the proposed method were observed to overlap with those obtained using the true attenuation map (TAAC method) (Fig. 3). Further, the AUC obtained by the proposed method was statistically similar to that obtained with the TAAC method (Table 1). The overlapped ROC curves and similar AUCs indicate that the proposed method, without access to the attenuation map, yields similar performance on the perfusion defect detection task as was obtained using the true attenuation map. Additionally, it was observed that the proposed method significantly outperformed the UAAC method, i.e., the method that used a uniform attenuation map.

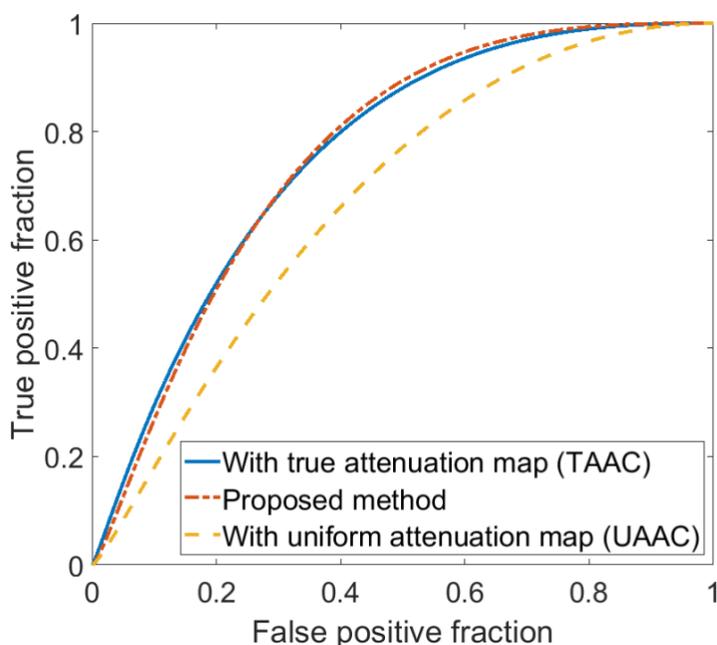

Fig. 3: ROC curves obtained using the proposed method, the TAAC method (true attenuation map), and the UAAC (uniform attenuation map) method. The ROC curves of the proposed method and the TAAC method were overlapped. Further, the proposed method outperformed the UAAC method.

| AC Method | Proposed method | TAAC method | UAAC method |
|---|---|---|---|
| **AUC and confidence intervals** | 0.76 [0.74, 0.78] | 0.76 [0.73, 0.78] | 0.68 [0.65, 0.70] |

Table 1: AUC obtained using the proposed method, the TAAC method (using true attenuation map), and the UAAC method (using uniform attenuation map). The values in brackets denote the 95% confidence intervals. The AUC obtained by the proposed method and the TAAC method were statistically similar.

### 3.2 Representative visual results

Fig. 4 shows some reconstructed activity maps using the TAAC and the proposed method. We observed that the images generated using the proposed method visually look similar to those when the true attenuation map was used.

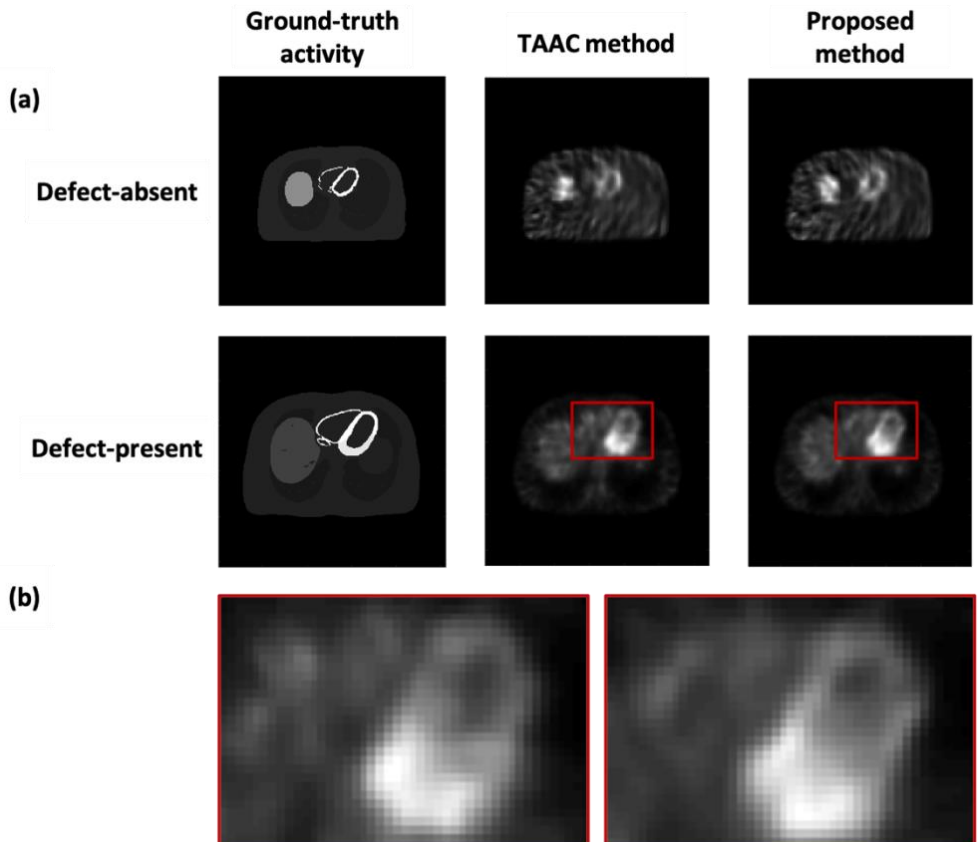

Fig. 4: (a) Representative reconstructed activity images examples using the true attenuation map (middle column) and the proposed method (right column). The upper example was a defect-absent case, while the lower sample was a defect-present case. (b) Zoomed version of the red-boxed regions in (a). Our results show that images generated using the proposed approach were visually similar to those obtained with the TAAC approach.

Overall, these results indicate that the proposed method yielded attenuation maps using only the SPECT emission data that yield similar task performance on a detection task as that obtained with true attenuation map and similar reconstructed activity images.

## 4 CONCLUSION AND DISCUSSIONS

We developed a physics and learning-based transmission-less attenuation compensation method for SPECT. This method uses the SPECT emission data acquired in the scatter-energy windows to estimate the attenuation distribution. A virtual imaging trial was performed to evaluate the performance of the method on the clinical task of detecting perfusion defects from SPECT images reconstructed using the proposed approach. Our results showed that the proposed method yielded task performance that was equivalent to that obtained when the true attenuation map was used. Further, the reconstructed activity maps obtained using the two approaches were visually similar. Additionally, the proposed method outperformed a method that used uniform attenuation map.

The results motivate further evaluation of the method using physical-phantom and clinical studies. Promising results in these studies would indicate that the proposed method can yield reliable performance in clinical settings for AC in SPECT when a transmission scan, such as a CT scan, is unavailable. Another area of future research is using list-mode data to reconstruct the attenuation map. The list-mode format stores data in more precise format compared to binned data, and this has been seen to result in improved image quality[37]-[40]. In fact, a recent study has shown that list-mode data may contain more information to estimate attenuation distribution compared to binned data[17]. Thus, expanding this method to directly process list-mode data may lead to improved AC.

In this work, we evaluated our method in the context of myocardial perfusion SPECT. SPECT has multiple other clinical applications, such as in assessment of neurodegenerative disorders[31],[42], and in imaging-based dosimetry[43-45]. Our method is general and the results motivate developing and evaluating the proposed method for those applications.

Our method has some limitations. The attenuation coefficients assigned to segmented regions were pre-defined. Assuming known attenuation coefficients can lead to inaccuracy in regions, such as lungs where the density varies on several factors, including the stage of disease. Evaluation of the method with clinical data will provide insights on the sensitivity of the technique to these variations. A limitation of the evaluation study is that it was simulation based. While we attempted for these studies to be clinically realistic, we recognize that our simulations may not model aspects related to instrumentation and accounting for variabilities in patient anatomy and physiology. In particular, modeling patient variability is important when evaluating deep-learning-based methods using simulations. There are ongoing efforts on approaches to evaluate simulated images such that their distribution matches those of clinical images [46],[47], which will help provide further confidence in simulation-based validation studies. Validation of the proposed method with physical-phantom and clinical studies, as mentioned above, would also help address these limitations.

In conclusion, a physics and learning-based transmission-less attenuation compensation method yielded reliable performance as objectively evaluated using a virtual-imaging-trial-based framework conducted in the context of detecting perfusion defects in myocardial SPECT. The method yielded similar performance as that obtained when the attenuation map was available, providing evidence that the method could yield reliable attenuation compensation even without a transmission scan.

## ACKNOWLEDGEMENTS

This work was financially supported by NIH R21 EB024647 (Trailblazer award) and by an NVIDIA GPU grant. The authors also thank Paul Segars (paul.segars@duke.edu) and Duke University for providing the XCAT phantom.